# NON-INTRUSIVE WATER USAGE CLASSIFICATION CONSIDERING LIMITED TRAINING DATA

**Pavlos Pavlou[1], Stelios Vrachimis[1, 2],**
**Demetrios G. Eliades[1], Marios M. Polycarpou[1,2]**

[1]KIOS Research and Innovation Center of Excellence, University of Cyprus, Nicosia, Cyprus
[2] Department of Electrical and Computer Engineering, University of Cyprus, Cyprus

[1]0000-0001-8740-4187, pavlou.v.pavlos@ucy.ac.cy, [2]0000-0001-8862-5205, vrachimis.stelios@ucy.ac.cy,
[3]0000-0001-8862-5205, eldemet@ucy.ac.cy, [4]0000-0001-6495-9171, mpolycar@ucy.ac.cy

## Abstract

*Smart metering of domestic water consumption to continuously monitor the usage of different appliances has been shown to have an impact on people's behavior towards water conservation. However, the installation of multiple sensors to monitor each appliance currently has a high initial cost and as a result, monitoring consumption from different appliances using sensors is not cost-effective. To address this challenge, studies have focused on analyzing measurements of the total domestic consumption using Machine Learning (ML) methods, to disaggregate water usage into each appliance. Identifying which appliances are in use through ML is challenging since their operation may be overlapping, while specific appliances may operate with intermittent flow, making individual consumption events hard to distinguish. Moreover, ML approaches require large amounts of labeled input data to train their models, which are typically not available for a single household, while usage characteristics may vary in different regions. In this work, we initially propose a data model that generates synthetic time series based on regional water usage characteristics and resolution to overcome the need for a large training dataset with real labeled data. The method requires a small number of real labeled data from the studied region. Following this, we propose a new algorithm for classifying single and overlapping household water usage events, using the total domestic consumption measurements. The classification procedure is described below: 1) During the offline feature learning stage, a dataset of labeled data corresponding to water-use profile signals is analyzed to some predefined features, such as event volume, event duration, event flow peak, and event signature, to extract its statistical properties, 2) The event classification stage monitors the provided measurement time-series for events between zero-flow intervals. The identified events are then classified using Dynamic Time Wrapping and an optimization procedure that finds the best label for the observed event based on the features learned in the first stage and similarity indices. Non-classified events are processed using a variation vector technique to identify the combined events which are then split into sub-single events and classified.*

## 1    INTRODUCTION

The increasing water consumption due to population growth and excessive urban development is creating an unbalanced situation between water demand and supply [1]. Adding to this, the need for continuous water supply and sufficient pressure during peak times puts even more burden on the water utilities that must face these challenges [2]. Among others, water management practices have been proposed as a response to these problems, aiming to ensure water demand needs are met constantly while promoting water conservation [3].

New advancements in sensor technology for collecting, analyzing, and transmitting high-resolution data to both utilities and consumers, are considered important tools for water management [4]. Smart metering of domestic water consumption to continuously monitor the usage of different appliances has been shown to have an impact on people's behavior towards water conservation [5] and can be a useful tool for water utilities in managing demand during peak hours and drought periods thus eliminating the need for further investment in upgrading the water infrastructure [6].

Smart metering can be categorized into intrusive and non-intrusive metering. Intrusive metering considers the installation of a sensor in each water-consuming appliance (e.g., dishwasher, toilet, shower) while non-intrusive metering considers the installation of only one sensor on the main water supply pipe of a house thus measuring the total household consumption. Although intrusive metering offers more insight into consumer habits, the installation of multiple sensors to monitor each appliance may have a high initial cost and may be inapplicable due to practical considerations [7].

Real-time data that are available through new smart metering systems must be coupled with data analytic techniques and intelligent algorithms to play a significant role as a decision-making tool and to have an impact on water demand management and water conservation. Disaggregation algorithms process the data retrieved through non-intrusive metering and identify which water end-use appliance is active by analyzing the total water consumption signal. Identifying which appliances are in use through non-intrusive water usage classification is challenging since their operation may be overlapping while specific appliances may operate with intermittent flow making individual consumption events hard to distinguish.

Water end-use disaggregation belongs to the general spectrum of time series classification problems. Time series classification is extensively addressed using machine learning and deep learning methodologies which require large training datasets [8] as well as with pattern recognition techniques based on similarity measurements such as Dynamic Time Wrapping (DTW) [9] and Longest Common Subsequence (LCSS) [10] that generally require a reference dataset. Various studies have been conducted to address the challenge of water end-use classification using smart water metering. In a first approach (Trace Wizard and Identiflow), decision tree methods were applied for water end-use classification which required significant data [11,12]. In [13,14], the authors suggested the use of pressure sensors combined with a Bayesian approach to identify water usage events (Hydrosense). These approaches required a high initial cost for the deployment of the sensor network and did not achieve high accuracy. Non-intrusive metering combined with machine learning methods were further used to disaggregate water end-use events. The authors in [15] proposed the use of an adaptable neuro-fuzzy network to classify water end-uses achieving high accuracy, using a limited dataset of flow measurements. In more recent studies, machine learning and data analytic algorithms were developed to address the problem of water end-use disaggregation, with promising results [16–21]. Several drawbacks that were noted in these studies include the need for a large amount of historical data to train the model and the absence of disaggregation techniques for combined water events. A notable study by [22] (Autoflow) addressed the aforementioned drawbacks using a hybrid combination of Hidden Markov Models, Artificial Neural Networks, and DTW algorithms, which was further improved to avoid the need of collecting new use-data for different regional use cases [23]. The "Autoflow" model addresses the classification of single and combined water end-use events with 85.9-96.1% and 81.8-91.5% accuracy respectively. However, as stressed by the authors, more regional data are needed to improve the performance of this method.

This work has two main contributions:

- Proposes an approach for calibrating an existing synthetic time-series data generator based on regional water usage characteristics and resolution. The generated data can be used to train Machine Learning algorithms without the need of collecting real labeled data for long periods from pilot studies.

- The development of a new methodology for classifying single and overlapping household water usage events within the same dataset using non-intrusive metering. The proposed approach takes into consideration water end-use events which exhibit intermittent or non-uniform flow.

The paper is structured as follows: Section 2 describes the data models, Section 3 provides the proposed classification methodology, Section 4 presents the performance of our classification approach and in Section 5 we conclude the paper and discuss some future extensions.

## 2     DATA MODEL

### 2.1     Available usage characteristics model

In this study, we use the available usage characteristics incorporated in the *STochastic Residential water End-use Model* (STREaM) introduced by [24]. STREaM is a modeling software that generates synthetic time series of data of a household with up to 10s resolution and it was calibrated on a large dataset including observed and disaggregated water end-uses from more than 300 single-family households in nine U.S. cities [25]. Each of the water end-uses considered in the STREaM dataset (toilet, shower, faucet, clothes washer, dishwasher) is characterized by its signature (i.e., typical consumption pattern) and the probability distributions of the water event volume, the single-use durations, the number of uses per day and the time of use during the day. The number of events per day is modelled using the negative binomial and Poisson distributions, the event start time with the Kernel distribution, and the event volume and duration with two-component Gaussian Mixtures. The probability distributions are created by taking into consideration the number of house residents and the efficiency of each appliance (standard or high efficiency.

The STREaM data model requires as inputs the number of household occupants, the available water appliances with their corresponding efficiency level, the simulation time, and the data resolution. Following, it generates time series of each water end-use and their sum as the total household water consumption based on the following procedure: i) samples the number of events for each water end-use and each day of the simulation time using the Monte-Carlo method from its probability distributions, ii) samples using the Monte-Carlo method the event-usage characteristics, duration, volume and time of use form their probability distributions, iii) randomly chooses one of the available signatures of the selected water end-use, iv) scales the duration and magnitude of the signature to match the pre-selected event duration and volume, and v) positions the newly created event time-series in the total event time series of the selected water end-use according to its start time.

### 2.2     Model calibration using limited regional data

The data model proposed in this work extends the STREaM data model to generate synthetic data based on regional water usage characteristics. For this, we assume that we have water usage data from a limited number of households within the region. We use a 1-week dataset from a single-family house in Cyprus to update the existing signatures and generate data with up to 1s resolution. The regional dataset includes data from the following appliances: toilet, shower, faucet, clothes washer and dishwasher. We assume that these data have been correctly classified per their usage and were collected at a resolution of 1s. Finally, it is assumed that no leakages exist in the recorded data.

The drawback of having a small dataset is that we may not be able to identify the probability distribution describing event occurrence, volume, and duration. However, the characteristic signatures of events can be identified even from this small dataset which are more representative of the appliances and local usage characteristics. Thus, the main approach for the development of the data model relies on updating the existing signatures with regional signatures from the case study. In addition, during the last step of the event generation process which includes the scaling of the duration and magnitude of the selected signature, boundaries were applied to ensure that generated events comply with the consumption flow rate indicated by the regional signatures. For example, during the sampling process of the usage characteristics, our model could pick a water end-use with a short-time duration and large volume resulting in an event with an inconsistent consumption pattern compared to the regional signature.

In the following paragraphs, we describe the methodology for the creation of regional consumption patterns. Signatures from the regional labeled dataset are extracted using a hybrid combination of DTW algorithm, k-medoids clustering method evaluated based on the "Silhouette index" and an affinity search technique. We use DTW in a clustering procedure to extract water end-use signatures from the regional dataset. The partitioning algorithm k-medoids splits the time series dataset into *k* clusters based on the minimum distance between the points of a cluster and a specified point at the center of the cluster and can be considered faster than other clustering methods [26]. The silhouette method measures the consistency of each cluster by comparing the similarity of an object to its cluster, compared to the remaining clusters of the group [27]. Silhouette score ranges from -1 to +1, with high values indicating a better fit of the object to its predefined cluster.

Initially, the time series (events) of each water appliance are extracted from the dataset, pre-processed to remove potentially faulty sensor measurements, and normalized to avoid scale differences. A similarity matrix for each group of events is obtained using DTW followed by k-Medoids clustering. The "Silhouette index" is used to define the number of clusters per fixture and the prototype signature is generated using a similarity search technique.

Each event time-series $S_t = \{S_1, \ldots, S_n\}$ comprised of *n* flow data points, is normalized to have a zero mean and standard deviation of one, thus being invariant to scale and offset, as follows:

$$\bar{S}_t(t) = \frac{S_t(t) - m}{\sigma} \qquad (1)$$

where the arithmetic mean *m* is given by:

$$m = \frac{\sum_{t=1}^{n} S_t}{n} \qquad (2)$$

and the standard deviation is given by:

$$\sigma = \sqrt{\frac{\sum_{t=1}^{n}(S_t - m)^2}{n}} \qquad (3)$$

The similarities between the time series of each group of events are calculated using the DTW method, resulting in the similarity matrix M of size AxA, where *A* the number of events per water-end use category, and the matrix elements are calculated as follows:

$$M_{ij} = W(\bar{S}_i, \bar{S}_j) \qquad (4)$$

where function $W(\cdot)$ calculates the distance between points of two time-series, using the DTW method. DTW is a methodology to measure the shape similarity between two time-series with different lengths. DTW wraps the time axis to align the data points and calculates the optimal alignment between two time-series according to the following equation:

$$W(i,j) = w(i,j) + min\{W(i-1,j), W(i-1,j-1), W(i,j-1)\} \tag{5}$$

where $s = \{s_1, .., s_i, .., s_m\}$ and $t = \{t_1, .., t_j, .., t_n\}$ are the two time series with $m$ and $n$ data points, respectively. Distance metric $w$ is given by $w(i,j) = |s_i - t_j|$ with the possible combinations limited to $(i-1,j)$, $(i-1,j-1)$, $(i,j-1)$. The accumulated DTW distance $W(m,n)$ is considered the optimal alignment between the two time-series, with initial condition $W(1,1) = w(1,1)$.

Following, the time series of each water fixture are grouped into clusters based on their similarity using the k-medoids clustering approach. Since the k-medoids method requires the number of clusters to be defined prior to clustering, the process can be carried out for a given range of clusters (e.g., 2-10 clusters). In order to define the appropriate number of clusters per water fixture, an evaluation method was simultaneously applied using the "Silhouette index".

The last step includes the extraction of the most representative signature of each cluster according to the DTW similarity results [28]. The time series with the lowest total dissimilarity $M_{xy}$ is extracted as the main signature:

$$d_n(S_x) = \left(\frac{\sum_{y \in C_n} M_{xy}}{TC_n}\right), D_n = \min(d_n) \tag{6}$$

where:

$M_{xy}$ : similarity matrix bewteen time series $S_x$ and $S_y$ belonging to cluster $C_n$

$TC_n$ : number of time series in each cluster $C_n$

$n$ : number of clusters

The extracted signature from each water end-use category can be eventually smoothened using polynomial fitting to remove measurement noise caused by the sensor and then stored in the data model. To illustrate the approach, we utilize a water-use dataset collected from a single-family household in Cyprus, in which water consumption was recorded with a 1-second resolution, and the data were labeled as toilet, shower, faucet, clothes washer, and dishwasher. Figure 1 shows the signatures extracted from each cluster of the shower category.

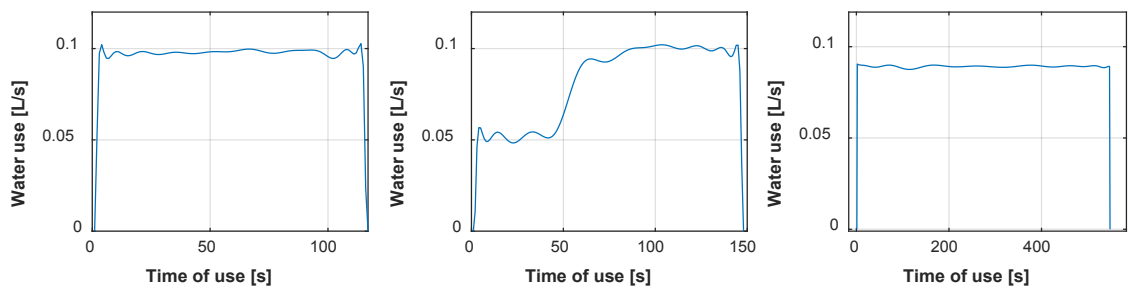

*Figure 1: Signature patterns for shower water end-use*

## 2.3   Datasets

Two synthetic datasets with a duration of 45 and 15 days respectively and 1s resolution were produced from the data model considering the following water end-uses: standard toilet, standard shower, standard faucet, high-efficiency clothes washer, and standard dishwasher. The 45-day dataset serves as the training set and the 15-day as the testing set. The training set is used to identify potential usage characteristics for each water end-use category and the testing set to evaluate the performance of the classification model described in the next section.

## 3 CLASSIFICATION METHODOLOGY

The water end-use classification procedure consists of two main stages.

1. In the first stage, namely *the offline feature learning stage,* the training dataset consisting of labeled data corresponding to water end-use signals is analyzed to extract the statistical properties of some predefined features including event duration, event volume, event flow peak, and event signature.

2. The *event classification stage monitors* the provided measurement time series from the test set for events between zero-flow intervals followed by the single and overlapping event classification. The classification of water end-use event relies on the DTW approach and an optimization procedure that uses similarity indices and statistical bounds extracted from the features learned in the first stage. Classification of events with an intermittent flow such as Dishwashers (DW) and Clothes washers (CW) are further processed considering a time window in the time series analysis that includes the device cycle in its entirety.

### 3.1 Offline Feature Learning Stage

This stage assumes the availability of inflow data of a residential household labeled according to which appliance is operating. In our case, the training dataset extracted from the data model is analyzed. The algorithm first creates event sets from labeled data by acquiring the observed time-series data with event labels and separating events, creating the set of events $E$. The events with the same label $l$ are then gathered, creating the subsets of events $E_l \subset E$. The features of event duration, event volume, and event flow peak were extracted. The next step includes the calculation of the 99% confidence intervals of each feature from each water end-use. The statistical analysis showed that sets of data have a skewed distribution, thus the proposed confidence intervals were obtained by filtering out the 1% most distant data points. This was achieved by calculating the absolute distance between each data point and the arithmetic mean of the dataset. We considered that only the generated training dataset is available for the classification model and not all the data that is stored in the data model.

### 3.2 Event Classification Stage

**Overview of the event classification process**

This stage distinguishes individual events in the time-series by filtering out data points separated by a zero-flow time interval. The event classification process is applied on the extracted events and consists of the single event classification and the combined event disaggregation and categorization. A combined event includes two or more single events with overlapping operations. Single event classification is performed using a hybrid approach that includes DTW algorithm and criteria based on similarity indices using statistical bounds extracted from the features at the previous stage. Following, the combined event disaggregation takes place using initially a filtering method to split the combined event into sub-events which are then processed through the single event classification procedure. Besides the difficulty in identifying both single and overlapping events another two obstacles that were identified during the process are:

- DW and CW devices have a working cycle that exhibits intermittent flow. Classification of such events was performed using a sliding time window of measurements.

- The existence of single events with a varying flow rate that occurs in rare circumstances can be easily misclassified as combined events. To overcome this problem, a filtered variation vector technique is applied in the combined event classification procedure to identify these events.

The overall classification process is presented in Figure 2.

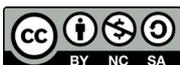

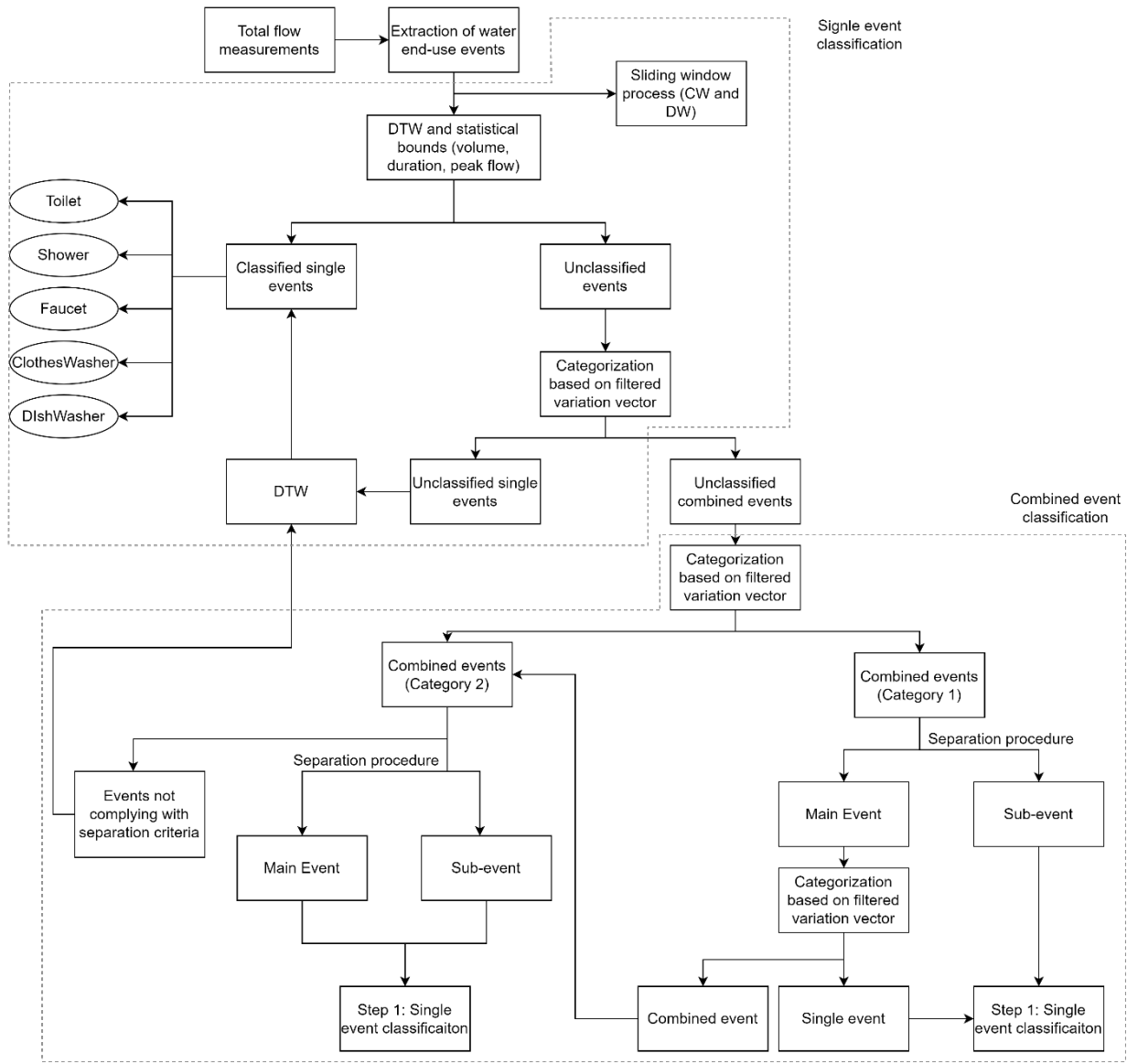

*Figure 2: Water end-use event classification process*

## Single event classification

The proposed single event classification relies mostly on pattern recognition through DTW. Initially, the investigated events and labeled signatures are normalized as described in equation 1. The first task is the detection of potential time windows inside the dataset with the operation of intermittent flow devices such as DW and CW. This is achieved by applying DTW between a sliding time window with a length equal to the full cycle of operation of the selected appliance and its corresponding labeled signatures. From the Cyprus case study pilot, in Figure 3a, the signature of a DW full-cycle operation is presented with a duration of 2793 seconds which corresponds to the time window used for the classification. Bounds of maximum flow criteria are also applied in this task to avoid misclassification of DW or CW time windows.

Following, DTW is applied in all events and distinguishes them into the following categories: toilet, shower, and faucet. Classification of WM and CW single events from their full cycle of operation is performed only within the time windows specified previously. In this case, the labeled signatures of WM and CW devices are broken down forming smaller sub-patterns (Figure 3b). A similarity matrix is created between the investigated event and the available labeled signatures stored in the database. Events with signature similarity above a specific threshold are then labeled. Simultaneously, a screening procedure is performed utilizing the minimum and maximum bounds obtained from features extracted from the training dataset (volume, duration, and peak flow). Any events not complying with the criteria defined through the DTW and the water end-use feature's statistical analysis are marked as unclassified.

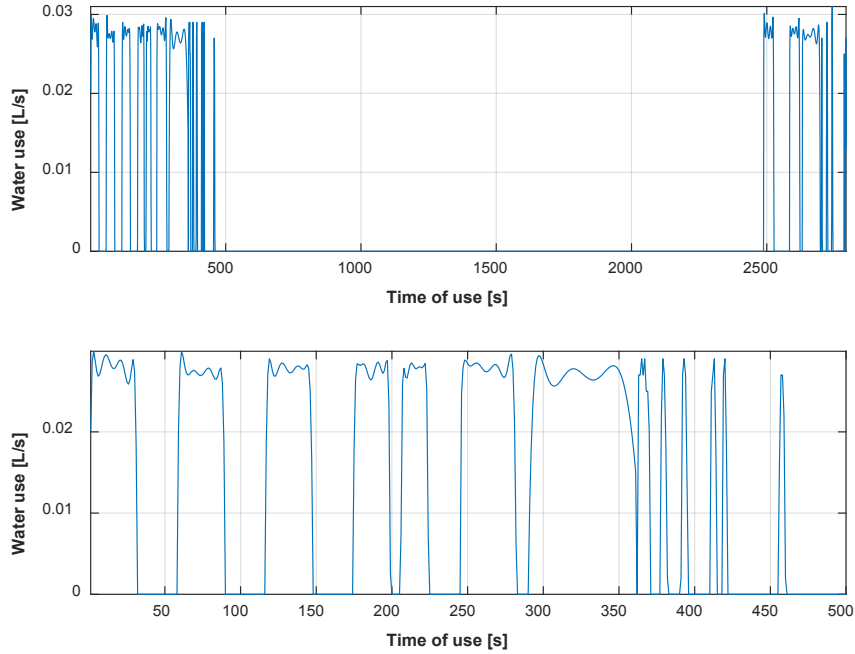

*Figure 3: a) Signature of Dishwasher's full operation cycle b) Sub-single events within the Dishwasher's main signature*

Unclassified events are then categorized into unclassified single and combined events. The categorization is performed using a filtering technique that detects flow rate changes within an event that exists at a specific threshold. Changes in the flow rates of an event are a good indication that another water-end use event has either been started or completed. The elements of the calculated vector are the differences between adjacent data points within an event, calculated as:

$$v_i = f_{i+1} - f_i, 1 \leq i < n \qquad (7)$$

Where $f = (f_1, f_2, .., f_i, .., f_n)$ the event flow rate points with a duration of $n$ seconds and $v = (v_1, v_2, .., v_i, .., v_{n-1})$ the extracted vector. A threshold is then specified to neglect fluctuations within the vector that do not correspond to the use of a new water appliance. A range of thresholds calculated based on the variation between the maximum flows of labeled events from the training dataset were evaluated and the value of 0.01 L/sec was selected as it achieved the highest accuracy. Unclassified single events are selected as the events which exhibit no fluctuations in the extracted filtered variation vector. The initial and final phases of the filtered vector are ignored since they mark the starting and ending of the event (Figure 4).

The main DTW classification methodology is applied again without using statistical bounds to categorize the unclassified single events. The remaining unlabeled events are considered as combined events and their classification follows in the next step.

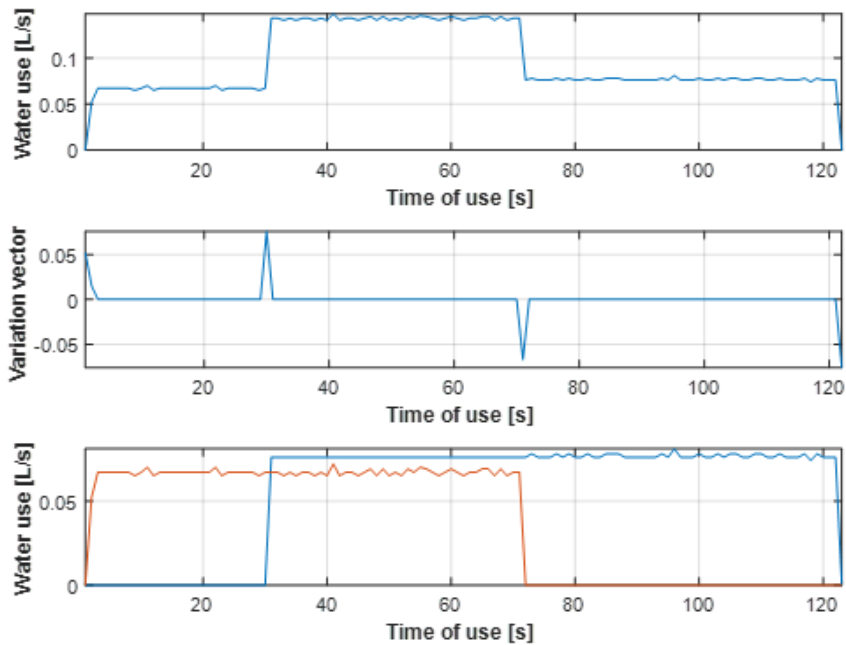

*Figure 4: a) Combined event as extracted from the dataset, b) Filtered variation vector of the combined event, c) Sub-events extracted from the original event*

**Combined event classification**

The combined event classification consists of two main tasks, the disaggregation of the combined event into single events and their classification following the approach described previously. Overlapping between events can be expressed in two different categories. The first category includes events overlapping with one sub-event a) starting and finishing before one or more other sub-events and b) starting and finishing after one or more other sub-events. The second category includes sub-events that start and finish within other sub-events.

The first step is the disaggregation of events belonging to the first category. This task is performed using an approach presented in [29], where the last flow-rate drop that corresponds to the finishing time of a combined event is compared to the last flow rate rise. If their difference is below a predefined threshold (a value of 0.005 L/sec resulted in the highest accuracy between a range of thresholds) then it is considered that a single sub-event occurred in the last phase of the combined event. The same principle applies to the starting phase. The sub-event is extracted from the initial combined event and the algorithm calculates its flow rate for the period that it was overlapping with other events. This is achieved by calculating the median flow rate during the period when only the targeted sub-event was active. An example is shown in Figure 4 with a sub-event starting and ending before the second sub-event. The remaining sub-event is evaluated again using the filtered variation vector approach and categorized as a single or combined event. If identified as a combined event, then it is included in the second category and processed as follows.

The second step includes the disaggregation of combined events included in the second category using the filtered variation vector defined previously. In this case, the algorithm searches within the filtered vector to identify the positions where a zero value is followed by a positive value and the positions where a negative value is followed by a zero value. These positions indicate the beginning and finishing of a sub-event within the combined event. The first "starting" position is matched with the first "finishing" position and the sub-event is separated from the base combined event. Events included in this category that do not meet these conditions (including at least one "starting" and "finishing" point) but they do present considerable fluctuations in their flow rate, are considered as single events and are then processed to the single event classification procedure with the use only of the DTW method. With this technique, single events with a varying flow rate that can be presented in real datasets (Figure 5), can be distiguinshed from combined events.

In the third step, the classification of the sub-events and the left-over (the remaining event after the separation process) base combined event extracted from the two previous steps takes place using the single event classification. Any events not classified are processed again through the combined event classification procedure.

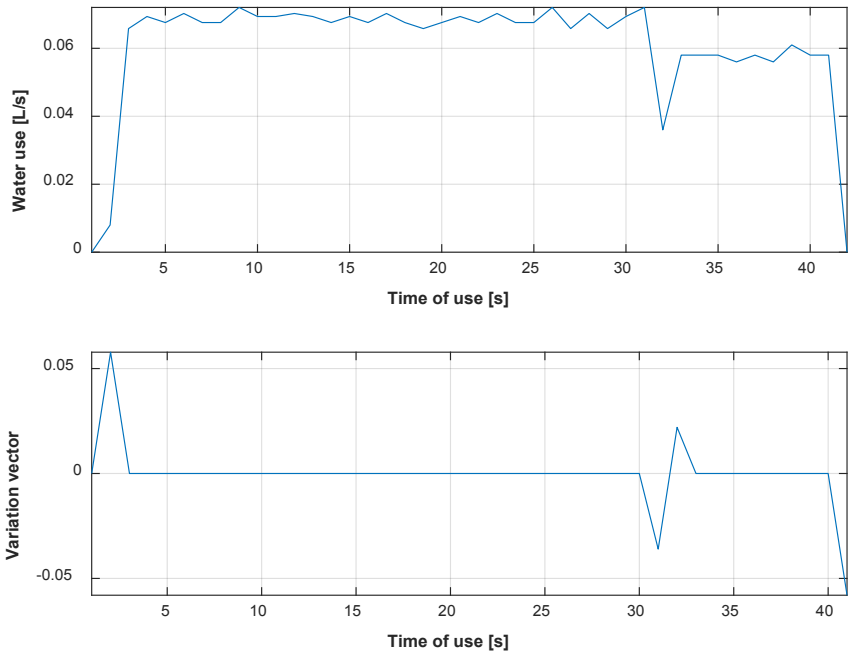

*Figure 5: Example of a single event initially misclassified as a combined event. a) Original event as extracted from the dataset, b) Filtered variation vector of the event*

## 4    RESULTS

**Evaluation metrics**

The macro f1-Score [30], a widely accepted metric that takes into consideration both the algorithm's precision and recall, is used:

$$\text{Macro f1-score} = \frac{2 \times precision \times recall}{precision + recall} \tag{8}$$

*Precision* indicates the percentage of true positive indices among the total number of positive indices classified by the model:

$$precision = TP/(TP + FP) \tag{9}$$

and *recall* measures the amount of correctly labeled positive cases among the total number of positive cases:

$$recall = TP/(TP + FN) \qquad (10)$$

*TP, TN, FP, FN* correspond to the number of true positives, true negative, false positive, and false negative events. The combination of the model's precision and recall makes F1-score less sensitive to imbalance classification scenarios and reaches its best value at 1 and worst score at 0. Testing accuracy is presented in terms of the number of events and consumption volume.

A confusion matrix is used to visually present the algorithm's performance by illustrating the number of correctly predicted events against the actual number of events.

**Confidence intervals**

The 99% confidence intervals were calculated from the statistical analysis of the three predefined features extracted from the training set (Table 1). For the DW and CW devices, the statistical analysis refers to the sub-single events that comprise a full cycle of operation. Toilet, faucet, and CW events have similar event characteristics, specifically for consumption duration and peak flow. Similarly, the calculated event volume bounds are identical as well, although CW can generate lower volume events than toilets and faucets. On the other hand, shower and DW events have more distinctive characteristics than the other categories which play a significant role in the classification process. Shower events have a longer duration, larger consumption volume, and a maximum flow higher than other categories. DW operation on the other side results in small events with low consumption and the lowest peak flow that can easily be distinguished from other appliances.

*Table 1: 99% confidence intervals obtained for the water end-use features: volume, duration, peak flow*

|                | Toilet    | Shower   | Faucet    | CW        | DW         |
|----------------|-----------|----------|-----------|-----------|------------|
| Duration (s)   | 10-190    | 90-880   | 10-170    | 1-139     | 1-85       |
| Volume (L)     | 0.66-9    | 13-90    | 0.43-10   | 0.03-11.85| 0.002-2.22 |
| Peak flow (L/s)| 0.04-0.10 | 0.09-0.15| 0.02-0.11 | 0.06-0.13 | 0.004-0.03 |

**Classification results**

The test set comprised of 1323 single and 22 combined events for a period of 15 days. The proposed approach has shown high accuracy (99%) in distinguishing the single events from the set of events while a lower F1-score of 69% was achieved for the combined event categorization although 77% of the combined events were correctly classified (Table 2). This is explained due to the existence of single events with a varying flow rate which were misclassified as combined events thus reducing the algorithm's precision. The calibration of the data model, which includes a large database of volume and duration features with regional water-end use signatures resulted in the development of a realistic dataset that included a few events with non-uniform consumption patterns. It was decided to keep these events in the dataset since they can indeed be presented in real conditions. An example is presented in Figure 5, showing a faucet event with an irregular flow trace. Although this event is considered rare, it is very realistic since it can be presented during the use of a single faucet (e.g during plate washing).

*Table 2: Accuracy results in distinguishing single and combined events*

|            | Single Events | Combined Events |
|------------|---------------|-----------------|
| Recall (%) | 99.2          | 77.3            |
| Precision (%) | 99.6 | 63.0 |
| Macro f1-score (%) | 99.4 | 69.4 |

**Single events**

Table 3 presents the results from the classification of single events in terms of the number of events and event volume. Scoring ranges from 83% to 98% in terms of the number of events and 84% to 99% in terms of volume. Single event classification precision is also presented through the confusion matrix (Figure 6) among the percentage of misclassified events per category.

**Toilet**: The model demonstrates an accuracy of 84% in classifying toilet events with 87% of the total toilet events being identified. In terms of volume, we notice a total score of 90% with approximately 91% of the total water volume consumed to be correctly calculated. Toilet events were mainly distinguished from the rest of the events due to their fixed mechanical operation/signature which was identified by the DTW algorithm. A few toilet events were misclassified with faucet events as presented in the confusion matrix due to similarity between their usage characteristics.

**Shower**: The highest recall score in terms of the number of events and volume was achieved for the shower appliance (100%) mainly due to its distinctive consumption volume, duration, and pattern characteristics. This score indicates that all shower events were correctly classified. The precision regarding the number of events, in this case, is lower (77%) though due to misclassification with faucet events. This occurs due to the presence of a small number of shower events with a short duration. This misclassification is not considered a limitation since the algorithm precision in terms of volume is considerably high (91%). The overall score for this category reaches 87% and 95% accuracy in terms of the number of events and volume, respectively.

**Faucet**: An 83% accuracy was achieved for faucet event classification with an 81% recall score regarding the total number of classified faucet events and 79% recall score for their corresponding volume. The lower score in terms of volume is explained by the misclassification of some single events as combined. As previously explained, a small number of single faucet events were misclassified as combined events due to their flow trace variation. Although in small number, these events had a considerably larger volume than typical faucet events which explained the variation between the two scoring categories.

**Clothes washer**: The model has also been able to correctly classify most of the CW events with 91% accuracy. The few misclassified events were confused with faucet events. The high score indicates the effectiveness of applying a sliding window to detect the full operation cycle of intermittent flow devices.

**Dishwasher**: Regarding the DW category, the model demonstrates the highest accuracy for both scoring categories (98-99%). Approximately all DW events were identified with the corresponding algorithm precision reaching 100%. The distinctive usage characteristics of DW events obtained from the statistical analysis along with the application of DTW using sliding windows proved to be highly efficient in detecting such events.

*Table 3: Single event classification accuracy in terms of number of events and volume*

| Number of events / Volume | Toilet | Shower | Faucet | CW | DW |
|---|---|---|---|---|---|
| Recall (%) | 86.7/91.4 | 100/100 | 81.4/78.7 | 91.3/91.0 | 95.7/98.7 |

| Precision (%) | 81.6/88.6 | 76.9/91.1 | 85.3/90.4 | 90.1/90.5 | 100/100 |
|---|---|---|---|---|---|
| **Macro f1-score (%)** | **84.1/90.0** | **87.0/95.3** | **83.3/84.2** | **90.7/90.8** | **97.8/99.4** |

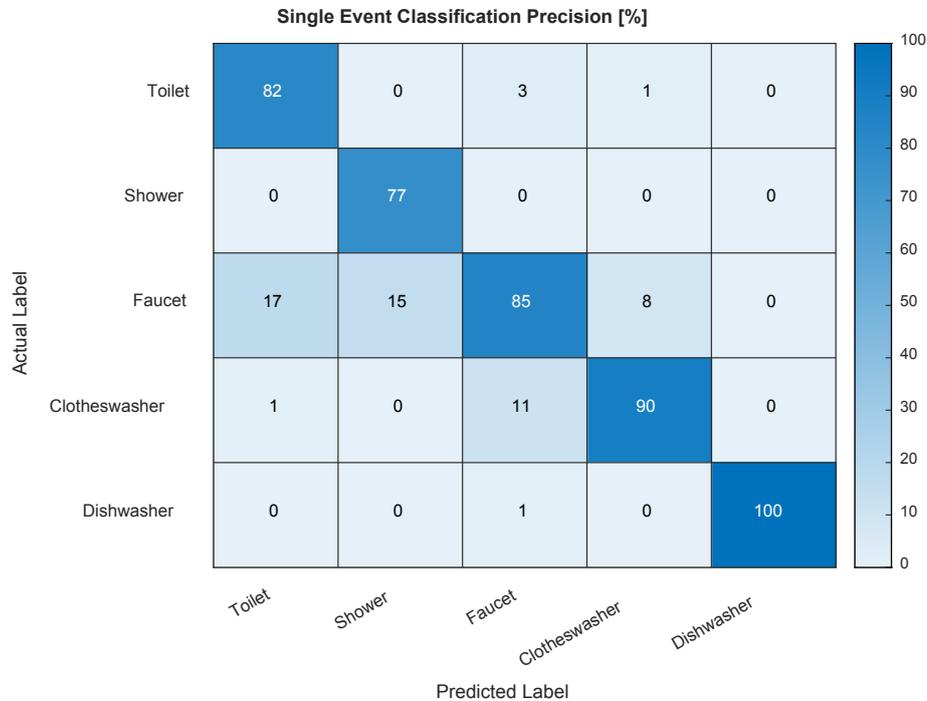

*Figure 6: Confusion matrix for single event classification precision (number of events)*

**Combined events**

As stated in Table 2, the algorithm correctly identified 17 out of 22 combined events (recall of 77%) using the proposed approach. The following approach consisting of the separation process and the classification of the extracted sub-events demonstrated an accuracy of 70%. Filtering out single events within combined events, which can be occurring completely at the same time or starting and finishing at the same time is considered a challenging task that needs to be further investigated. The extraction of sub-events under these circumstances is not always accurate, and the imbalance between the number of sub-events and single events can explain the lower classification score. Further improvements can be considered in the separation process to reach a higher precision of combined event separation and classification.

## 5  CONCLUSIONS AND FUTURE WORK

In this work, we initially presented an approach of extracting water end-use signatures from a limited real labeled dataset to calibrate our data model on regional water usage characteristics and resolution. The developed data model gives us the ability to use an existing large database of water end-use features from STREaM including event duration, volume, and number of events per day, and produce synthetic time series of events with regional consumption patterns. The method requires a small number of real labeled data from the target region. Following, a water end-use classification procedure is presented considering non-intrusive monitoring. The developed approach addresses the main difficulties of this challenging problem such as identifying overlapping events, devices with intermittent flow, and single events which exhibit a non-uniform consumption pattern. In the proposed hybrid approach, we use sliding windows, DTW, and confidence intervals to identify active water end-uses with accuracy ranging between 84-99% for single events and 70% for combined events. The main difficulties encountered were the identification of single events with varying flow rates and the accurate separation of combined events into sub-singe events. As shown in the results, the accurate extraction of single events from a combined event is crucial during the classification process. The applicability of this approach is further suggested to be tested in large real datasets from regions with different water usage characteristics considering also the presence of leakages.

# 6    ACKNOWLEDGMENTS

The work was supported by the FLOBIT Project EXCELLENCE/0918/0282 which is co-financed by the European Regional Development Fund and the Republic of Cyprus through the Research and Innovation Foundation, and the European Union Horizon 2020 program under Grant Agreement No. 739551 (KIOS CoE) and the Government of the Republic of Cyprus through the Deputy Ministry of Research, Innovation and Digital Policy.

# 7    REFERENCES

[1]    A. K. Biswas and C. Tortajada, "Assessing Global Water Megatrends," in *Water Resources Development and Management*, Springer, Singapore, 2018.

[2]    H. M. Ramos, A. Carravetta, and A. M. Nabola, "New Challenges in Water Systems," *Water*, vol. 12, no. 9, Aug. 2020, p. 2340.

[3]    A. Liu, D. Giurco, and P. Mukheibir, "Urban water conservation through customised water and end-use information," *Journal of Cleaner Production*, vol. 112, Jan. 2016, pp. 3164–3175.

[4]    A. Cominola, M. Giuliani, D. Piga, A. Castelletti, and A. E. Rizzoli, "Benefits and challenges of using smart meters for advancing residential water demand modeling and management: A review," *Environmental Modelling & Software*, vol. 72, Oct. 2015, pp. 198–214.

[5]    A. Cominola *et al.*, "Long-term water conservation is fostered by smart meter-based feedback and digital user engagement," *npj Clean Water*, vol. 4, no. 1, Dec. 2021, p. 29.

[6]    M. S. Rahim, K. Anh Nguyen, R. A. Stewart, D. Giurco, and M. Blumenstein, "Predicting Household Water Consumption Events: Towards a Personalised Recommender System to Encourage Water-conscious Behaviour," in *Proc. 2019 International Joint Conference on Neural Networks (IJCNN)*, Budapest, IEEE, 2019, pp. 1–8.

[7]    Y. Kim, T. Schmid, Z. M. Charbiwala, J. Friedman, and M. B. Srivastava, "NAWMS," in *Proc. of the 6th ACM conference on Embedded network sensor systems - SenSys '08*, 2008, p. 309.

[8]    A. P. Ruiz, M. Flynn, J. Large, M. Middlehurst, and A. Bagnall, "The great multivariate time series classification bake off: a review and experimental evaluation of recent algorithmic advances," *Data Mining and Knowledge Discovery*, vol. 35, no. 2, Mar. 2021, pp. 401–449.

[9]    E. Keogh and C. A. Ratanamahatana, "Exact indexing of dynamic time warping," *Knowledge and Information Systems*, vol. 7, no. 3, Mar. 2005, pp. 358–386.

[10]   M. Vlachos, G. Kollios, and D. Gunopulos, "Discovering similar multidimensional trajectories," in *Proc. 18th International Conference on Data Engineering*, San Jose, CA, IEEE, 2002, pp. 673–684.

[11]   W. B. DeOreo, J. P. Heaney, and P. W. Mayer, "Flow trace analysis to access water use," *Journal - American Water Works Association*, vol. 88, no. 1, Jan. 1996, pp. 79–90.

[12]   R. A. Stewart, R. Willis, D. Giurco, K. Panuwatwanich, and G. Capati, "Web-based knowledge management system: linking smart metering to the future of urban water planning," *Australian Planner*, vol. 47, no. 2, Jun. 2010, pp. 66–74.

[13]   J. E. Froehlich, E. Larson, T. Campbell, C. Haggerty, J. Fogarty, and S. N. Patel, "HydroSense," in *Proc. of the 11th international conference on Ubiquitous computing*, New York, ACM, 2009, pp. 235–244.

[14]   J. Froehlich *et al.*, "A Longitudinal Study of Pressure Sensing to Infer Real-World Water Usage Events in the Home," in *International conference on pervasive computing*, Springer, Berlin, 2011, pp. 50–69.

[15]   M. A. Corona-Nakamura, R. Ruelas, B. Ojeda-Magana, and D. Andina, "Classification of domestic water consumption using an Anfis model," in *Proc.* 2008 World Automation Congress, Waikoloa, HI, IEEE,

2008, pp. 1-9.

[16]    B. E. Meyer, K. Nguyen, C. D. Beal, H. E. Jacobs, and S. G. Buchberger, "Classifying Household Water Use Events into Indoor and Outdoor Use: Improving the Benefits of Basic Smart Meter Data Sets," *Journal of Water Resources Planning and Management*, vol. 147, no. 12, Dec. 2021, p. 04021079.

[17]    L. Pastor-Jabaloyes, F. Arregui, and R. Cobacho, "Water End Use Disaggregation Based on Soft Computing Techniques," *Water*, vol. 10, no. 1, Jan. 2018, p. 46.

[18]    R. Cardell-Oliver, J. Wang, and H. Gigney, "Smart Meter Analytics to Pinpoint Opportunities for Reducing Household Water Use," *Journal of Water Resources Planning and Management*, vol. 142, no. 6, Jun. 2016, p. 04016007.

[19]    J. S. Vitter and M. E. Webber, "A non-intrusive approach for classifying residential water events using coincident electricity data," *Environmental Modelling & Software*, vol. 100, Feb. 2018, pp. 302–313.

[20]    M. S. Rahim, K. A. Nguyen, R. A. Stewart, D. Giurco, and M. Blumenstein, "Machine Learning and Data Analytic Techniques in Digital Water Metering: A Review," *Water*, vol. 12, no. 1, Jan. 2020, p. 294.

[21]    F. Mazzoni, S. Alvisi, M. Franchini, M. Ferraris, and Z. Kapelan, "Automated Household Water End-Use Disaggregation through Rule-Based Methodology," *Journal of Water Resources Planning and Management*, vol. 147, no. 6, Jun. 2021, p. 04021024.

[22]    K. A. Nguyen, R. A. Stewart, H. Zhang, and C. Jones, "Intelligent autonomous system for residential water end use classification: Autoflow," *Applied Soft Computing*, vol. 31, Jun. 2015, pp. 118–131.

[23]    K. A. Nguyen, R. A. Stewart, H. Zhang, and O. Sahin, "An adaptive model for the autonomous monitoring and management of water end use," *Smart Water*, vol. 3, no. 1, Dec. 2018, p. 5.

[24]    A. Cominola, M. Giuliani, A. Castelletti, D. E. Rosenberg, and A. M. Abdallah, "Implications of data sampling resolution on water use simulation, end-use disaggregation, and demand management," *Environmental Modelling & Software*, vol. 102, Apr. 2018, pp. 199–212.

[25]    W. B. DeOreo, "Analysis of water use in new single family homes," *By Aquacraft. For Salt Lake City Corporation and US EPA*, 2011.

[26]    S. Aghabozorgi, A. Seyed Shirkhorshidi, and T. Ying Wah, "Time-series clustering – A decade review," *Information Systems*, vol. 53, Oct. 2015, pp. 16–38.

[27]    V. Vijayakumar and R. Bremananth, "A Study on Human Hair Analysis and Synthesis," in *Proc. International Conference on Computational Intelligence and Multimedia Applications (ICCIMA 2007)*, Sivakasi, IEEE, 2007, pp. 549–556.

[28]    S. Aghabozorgi, T. Ying Wah, T. Herawan, H. A. Jalab, M. A. Shaygan, and A. Jalali, "A Hybrid Algorithm for Clustering of Time Series Data Based on Affinity Search Technique," *The Scientific World Journal*, vol. 2014, 2014, pp. 1–12.

[29]    K. A. Nguyen, R. A. Stewart, and H. Zhang, "An intelligent pattern recognition model to automate the categorisation of residential water end-use events," *Environmental Modelling & Software*, vol. 47, Sep. 2013, pp. 108–127.

[30]    Haibo He and E. A. Garcia, "Learning from Imbalanced Data," *IEEE Transactions on Knowledge and Data Engineering*, vol. 21, no. 9, Sep. 2009, pp. 1263–1284.